\documentclass[a4paper,10pt,
twocolumn,colorlinks,urlcolor=black,linkcolor=black,citecolor=black,
]{article}
\usepackage{amsmath,amssymb}
\usepackage[dvips]{color}
\usepackage{cite}
\usepackage{hangcaption}
\usepackage{amsmath,amssymb}
\usepackage[dvips]{color}
\usepackage{cite}
\usepackage{eepic}
\usepackage{epic}
\usepackage{hangcaption}
\usepackage{amsmath}
\usepackage{amssymb}
\usepackage{amscd}
\usepackage{amsthm}
\usepackage[dvipdfmx,%
 bookmarks=true,%
 bookmarksnumbered=true,%
 colorlinks=true,%
 setpagesize=false,%
 pdfkeywords={TeX; dvipdfmx; hyperref; color;}]{hyperref}
\def\cases{\left\{\begin{array}{ll}}
\def\endcases{\end{array}\right.}

\def\bigtimes{\mathop{\mbox{\Large $\times$}}}
\begin{document}
\setcounter{page}{1}
\twocolumn
[
\vskip1.5cm
\begin{center}
{\LARGE \bf 
Zeno's Paradoxes in the Mechanical World View
}
\\
{$\;$}
\\
{\LARGE \bf
}
\vskip0.5cm
{\rm
\large
Shiro Ishikawa
}
\\
\vskip0.2cm
\rm
\it
Department of Mathematics, Faculty of Science and Technology,
Keio University, 
\\
3-14-1 Hiyoshi, Kohoku-ku, Yokohama, 223-8522 Japan ({E-Mail:
ishikawa@math.keio.ac.jp})
\end{center}
\par
\rm
\vskip0.3cm
\par
\noindent
\normalsize
\par
\noindent
There is a very reason to consider that to solve Zeno's paradoxes is to propose
the theory of mechanical world view. We believe that this is not only our 
opinion but also most philosophers' opinion. Recently, in order to justify 
Heisenberg`s uncertainty principle (cf. Rep. Math. Phys Vol. 29, No. 3, 1991)
more firmly. we proposed the linguistic interpretation of quantum mechanics
(called quantum and classical measurement theory), which was characterized as
the metaphysical and linguistic turn of the Copenhagen interpretation.
This turn from physics to language does not only extend quantum mechanics
to classical systems but also yield the (quantum and classical) mechanical
world view (and therefore, establish the method of science). If it be so,
we may assert that Zeno's paradoxes (Flying Arrow Paradox,  Achilles and the
tortoise, etc.) were already solved in measurement theory. The purpose of this
paper is to examine  this assertion.

\par
\par
\vskip0.1cm
\par
\noindent
\vskip1.0cm
\par
]

%
%

\par

\def\Cal{\cal}
\def\bigstimes{\text{\large $\: \boxtimes \,$}}

\def\LL{\lq\lq}
\def\RR{\rq\rq {$\;$}}

\par
\noindent
{
\begin{itemize}
\item[{\large \bf 1.}]
{
\large
\bf
Introduction}
\end{itemize}
}
\par
\noindent
{
\begin{itemize}
\item[{\bf 1.1.}]
{
\bf
Zeno's paradox (Achilles and the tortoise)}
\end{itemize}
}

Although there are several Zeno's paradoxes
({\it cf.} {{{}}}{\cite{Zeno}}),
we believe that they are essentially the same problem.
Thus, in this paper,
we devote ourselves to
the most famous Zeno's paradox
(i.e.,
Achilles and the tortoise
).
\begin{itemize}
\item[(A)]
{\sf [Achilles and the tortoise]}
In a race, the quickest runner can never overtake the slowest, 
since the pursuer must first reach the point whence the pursued started, 
so that the slower must always hold a lead. 
Is it true?
\end{itemize}
\par
\noindent
{
\begin{itemize}
\item[{\bf 1.2.}]
{
\bf
The formulation}
\end{itemize}
}
In what follows we shall introduce well known idea
about the above (A).
For example, assume that
the velocity 
$v_q$
[resp. $v_s$]
of the quickest [resp. slowest] runner
is equal to
$v(>0)$ [resp. ${\gamma}v \; ( 0<{\gamma}<1)$].
And further,
assume that
the position
of the quickest [resp. slowest] runner
at time $t=0$
is equal to
$0$
[resp. $a \; (>0)$].
Thus,
we can assume that
the position ${\xi (t)}$
of the quickest runner
and
the position
$\eta (t)$
of the slowest
runner
at time $t$
$( \ge 0)$
is respectively represented by
\begin{eqnarray}
\cases
\xi (t) =vt
\\
\eta(t) = {\gamma}vt + a
\endcases
\label{eq1}
\end{eqnarray}
\par
\noindent
\par
\noindent
{
\begin{itemize}
\item[{\bf 1.3.}]
{
\bf
Calculations}
\end{itemize}
}
\par
The formula {{{}}}{(\ref{eq1})} can be calculated as follows
(i.e.,
(i) or (ii)):
\par
\noindent
[(i): Algebraic calculation of {{{}}}{(\ref{eq1})}]:
\par
\noindent
Solving
$\xi(s_0)=\eta(s_0)$,
that is,
\begin{align*}
vs_0 = \gamma v s_0 + a
\end{align*}
we get
$s_0= \frac{a}{(1-\gamma) v}$.
That is,
at time
$s_0= \frac{a}{(1-\gamma) v}$,
the fast runner catches up with the slow runner.
\par
\noindent
[(ii): Iterative calculation of {{{}}}{(\ref{eq1})}]:
\par
\noindent
\par
\noindent
Define
$t_k$
$(k=0,1,...)$
such that,
$t_0=0$
and
\begin{align*}
t_{k+1}= \gamma v t_k + a
\;\;\;\;
(k=0,1,2,...)
\end{align*}
Thus,
we see that
$t_k=\frac{(1-{\gamma}^k)a}{(1-{\gamma})v}$
$(k=0,1,...)$.
Then,
we have that
\par
\noindent
\vskip-0.4cm
\par
\noindent
{
\begin{eqnarray}
\big(
\xi (t_k), \eta (t_k)
\big)
&
=
\big(
\frac{(1-{\gamma}^k)a}{1-{\gamma}},
\frac{(1-{\gamma}^{k+1})a}{1-{\gamma}}
\big)
\nonumber
\\
&
\to 
\big(
\frac{a}{1-{\gamma}},\frac{a}{1-{\gamma}}
\big)
\label{eq2}
\end{eqnarray}
}
\par
\noindent
\par
\noindent
as $k \to \infty $.
Therefore, the quickest runner catches up with the slowest
at time $s_0 =\frac{a}{(1-{\gamma})v}$.
\par
\noindent
[(iii): Conclusion]:
After all, by the above (i)
or (ii),
we can conclude that
\begin{itemize}
\item[(B)]
the quickest runner can overtake the slowest
at time
$s_0 =\frac{a}{(1-{\gamma})v}$.
\end{itemize}
\par
\noindent
\par
\noindent
\par
\noindent
{
\begin{itemize}
\item[{\bf 1.4.}]
{
\bf
What is Zeno's paradox?}
\end{itemize}
}
\par
\noindent

\par
It is a matter of course that
what is important is the formulation
(in Section 1.2)
and not
the calculation
(in Section 1.3).
That is,
we believe that
what Zeno 
(or,
the philosophers who inherit Zeno
)
wanted to ask
is
as follows:
\begin{itemize}
\item[(C)]
Why do we believe in the formula {{{}}}{(\ref{eq1})}?
$\;\;$
Or, what kind of world view
is the {{{}}}{(\ref{eq1})} based on?
\end{itemize}
The purpose of this paper is
to answer this question (C),
that is,
to assert that
\begin{itemize}
\item[(D)]
the formula {{{}}}{(\ref{eq1})} is based on measurement theory.
\end{itemize}
There is a very reason to consider that
the world view in (C) is called the mechanical world view.
Thus, if the assertion (D) is generally accepted, 
we may assert that
measurement theory
is just
the theory of mechanical world view.

\vskip0.2cm
\par
\noindent
{
\begin{itemize}
\item[{\large \bf 2.}]
{
\large
\bf
Measurement Theory
}
\end{itemize}
}
\par
\par
\noindent
\begin{itemize}
\item[{\bf 2.1.}]
\bf
Overview: Measurement theory
\end{itemize}

\rm
\par
\par
\noindent

In this section,
we shall mention the overview of measurement theory
(or in short, MT).
\par
\par
\rm

\par
\noindent
\par
It is well known ({\it cf.}{{{}}}{\cite{Neum}}) that
quantum mechanics is
formulated 
in an operator algebra $B(H)$
(i.e.,
an operator algebra
composed of all
bounded linear operators
on a Hilbert space $H$
with the norm
$\|F\|_{B(H)}=\sup_{\|u\|_H = 1} \|Fu\|_H$
)
as follows:
\begin{itemize}
\item[(E)]
$
\quad
\underset{\text{\scriptsize (physics)}}{\text{quantum mechanics}}
\qquad
$
\item[]
$
=
{}
{}
\displaystyle{
{
\mathop{\mbox{[quantum measurement]}}_{\text{\scriptsize (probabilistic interpretation) }}
}
}
{}
{}
+
{}
{}
\displaystyle{
\mathop{
\mbox{
[causality]
}
}_{
{
\mbox{
\scriptsize
(
kinetic
equation)
}
}
}
}
{}
$
\end{itemize}
\par
\noindent
Also, the Copenhagen interpretation due to N. Bohr (et al.) is characterized
as
the guide to the usage of quantum mechanics (E).
Although
quantum mechanics (E) with the Copenhagen interpretation
is generally accepted as one of the most
trustworthy theories
in science,
it should be noted that
there is no definitive statement of the Copenhagen interpretation,
that is,
there are a number of ideas that are associated with the Copenhagen interpretation.
We do not think that this fact is desirable.

\par
Measurement theory
(mentioned in the following sections
or refs.
{{{}}}{\cite{Ishi1,Ishi2,Ishi3,Ishi4,Ishi5,Ishi6,Ishi7,Ishi8,Ishi9,Ishi10,Ishi11}})
is,
by an analogy of the (E), constructed
as the mathematical
theory
formulated
in a certain $C^*$-algebra ${\cal A}$
(i.e.,
a norm closed subalgebra in $B(H)$,
{\it cf.} {{{}}}{\cite{Saka}}
)
as follows:
\par
\par
\begin{itemize}
\item[(F)]
$
\quad
\underset{\text{\scriptsize (
language)}}{\text{measurement theory}}
\!
$
\item[]
$
=
\displaystyle{
{
\mathop{\mbox{[measurement]}}_{\text{\scriptsize (Axiom 1 in Section 2.2) }}
}
}
+
\displaystyle{
\mathop{
\mbox{
[causality]
}
}_{
{
\mbox{
\scriptsize
(Axiom 2 in Section 2.2)
}
}
}
}
$
\end{itemize}
\par
\par
\noindent
Note that this theory (F) is not physics but a kind of language
based on
{\lq\lq}the mechanical world view{\rq\rq}
since it is a mathematical generalization of 
quantum mechanics (E).

When ${\cal A}=B_c(H)$,
the ${C^*}$-algebra composed
of all compact operators on a Hilbert space $H$,
the (F) is called {quantum measurement theory}
(or,
quantum system theory),
which can be regarded as
the linguistic aspect of quantum mechanics.
Also, when ${\cal A}$ is commutative
(that is, 
when ${\cal A}$ is characterized by $C_0(\Omega)$,
the $C^*$-algebra composed of all continuous 
complex-valued functions vanishing at infinity
on a locally compact Hausdorff space $\Omega$
({\it cf.} {{{}}}{\cite{Yosi, Saka}})),
the (F) is called {classical measurement theory}.
Thus, we have the following classification:
\begin{itemize}
\item[(G)]
$
\quad
\underset{\text{\scriptsize }}{\text{measurement theory}}
$
\\
$
\;
=
\left\{\begin{array}{ll}
\underset{\text{\scriptsize (when ${\cal A}=B_c (H)$)}
}{\text{quantum measurement theory}}
\\
\underset{\text{\scriptsize (when ${\cal A}=C_0(\Omega)$)}}{\text{classical measurement theory}}
\end{array}\right.
$
\end{itemize}
That is, this theory covers 
several conventional system theories
(i.e., statistics, dynamical system theory,
quantum system theory).


\par
\noindent
\begin{itemize}
\item[{\bf 2.2.}]
\bf
Measurement theory
\end{itemize}

\par
\noindent
\par
Measurement theory (F) has two formulations
(i.e.,
the $C^*$-algebraic formulation
and
the $W^*$-algebraic formulation,
{\it cf.} {{{}}}{\cite{Ishi4}}
).
In this paper,
we devote ourselves to
the $W^*$-algebraic formulation.
of the measurement theory
(F).

Let
${\cal A}
( \subseteq B(H))$
be
a
${C^*}$-algebra,
and let
${\cal A}^*$ be the
dual Banach space of
${\cal A}$.
$\;\;$
That is,
$ {\cal A }^* $
$ {=}  $
$ \{ \rho \; |$
$ \; \rho$
is a continuous linear functional on ${\cal A}$
$\}$,
and
the norm $\| \rho \|_{ {\cal A }^* } $
is defined by
$ \sup \{ | \rho ({}F{}) |  \:{}\; | \;\; F \in {\cal A}
\text{ such that }\| F \|_{{\cal A}} 
(=\| F \|_{B(H)} )\le 1 \}$.
Define the
\it
mixed state
$\rho \;(\in{\cal A}^*)$
\rm
such that
$\| \rho \|_{{\cal A}^* } =1$
and
$
\rho ({}F) \ge 0
$
for all
$F\in {\cal A}$
such that
$
F \ge 0$.
And define the mixed state space
${\frak S}^m  ({}{\cal A}^*{})$
such that
\begin{align*} {\frak S}^m  ({}{\cal A}^*{})
{=}
\{ \rho \in {\cal A}^*  \; | \;
\rho
\text{ is a mixed state}
\}.
\end{align*}
%
\rm
A mixed state
$\rho (\in {\frak S}^m  ({\cal A}^*) $)
is called a
\it
pure state
\rm
if
it satisfies that
{\lq\lq $\rho = \theta \rho_1 + ({}1 - \theta{}) \rho_2$
for some
$ \rho_1 , \rho_2 \in {\frak S}^m  ({\cal A}^*)$
and
$0 < \theta < 1 $\rq\rq}
implies
{\lq\lq $\rho =  \rho_1 = \rho_2$\rq\rq}\!.
Put
\begin{align*} {\frak S}^p  ({}{\cal A}^*{})
{=}
\{ \rho \in {\frak S}^m  ({\cal A}^*)  \; | \;
\rho
\text{ is a pure state}
\},
\end{align*}
which is called a
\it
state space.
\rm
\rm
It is well known
({\it cf.} {{{}}}{\cite{Saka}})
that
$ {\frak S}^p  ({}{B_c(H)}^*{})=$
$\{ | u \rangle \! \langle u |
$
(i.e., the Dirac notation)
$
\:\;|\;\:
$
$
\|u \|_H=1 
\}$,
and
$ {\frak S}^p  ({}{C_0(\Omega)}^*{})$
$=$
$\{ \delta_{\omega_0} \;|\; \delta_{\omega_0}$ is a point measure at
${\omega_0}
\in \Omega
\}$,
where
$ 
\int_\Omega f(\omega) \delta_{\omega_0} (d \omega )$
$=$
$f({\omega_0})$
$
(\forall f
$
$
\in C_0(\Omega))$.
The latter implies that
$ {\frak S}^p  ({}{C_0(\Omega)}^*{})$
can be also identified with
$\Omega$
(called a {\it spectrum space}
or simply
{\it spectrum})
such as
\begin{align}
\underset{\text{\scriptsize (state space)}}{{\frak S}^p  ({}{C_0(\Omega)}^*{})}
\ni \delta_{\omega} \leftrightarrow {\omega} \in 
\underset{\text{\scriptsize (spectrum)}}{\Omega}
\label{eq3}
\end{align}

\par
\vskip0.2cm
\par
\rm

\par
\par
Consider the pair
$[{\cal A},{\cal N}]_{B(H)}$,
called a
{\it
basic structure}.
Here,
${\cal A} ( \subseteq B(H))$
is a $C^*$-algebra,
and
${\cal N}$
(${\cal A} \subseteq {\cal N} \subseteq B(H)$)
is a particular $C^*$-algebra
(called a $W^*$-algebra)
such that
${\cal N}$ is the weak closure of
${\cal A}$
in $B(H)$.
Let
${\cal N}_*$
be the pre-dual Banach space.

For example,
we see ({\it cf.} {{{}}}{\cite{Saka}}) that,
when ${\cal A}=B_c(H)$,
\begin{itemize}
\item[(i)]
${\cal A}^*=${\lq\lq}trace class",
${\cal N}=B(H)$,
${\cal N}_*=${\lq\lq}trace class".
\end{itemize}
Also, when ${\cal A}=C_0(\Omega)$,
\begin{itemize}
\item[(ii)]
${\cal A}^*=${\lq\lq}the space of all signed measures on $\Omega$",
${\cal N}=L^\infty ( \Omega, \nu)
(\subseteq B(L^2 ( \Omega, \nu)))$,
${\cal N}_*=L^1 ( \Omega, \nu)$,
where $\nu$ is some measure on $\Omega$
({\it cf.} {{{}}}{\cite{Saka}}).
\end{itemize}
In this paper,
$L^\infty ( \Omega, \nu)$
and
$L^1 ( \Omega, \nu)$
is often written by
$L^\infty ( \Omega)$
and
$L^1 ( \Omega)$
respectively.

For instance,
in the above (ii) we must clarify the meaning
of the {\lq\lq}value"
of
$F(\omega_0)$
for
$F \in L^\infty(\Omega, \nu )$
and
$\omega_0 \in \Omega$.
An element
$F (\in {\cal N} )$
is
said to be
{\it
essentially continuous
at
}
$\rho_0 (\in  {\frak S}^p  ({}{\cal A}^*{}))$,
if there uniquely exists a complex number $\alpha$
such that
\begin{itemize}
\item[(H)]
if $\rho$
$( \in {\cal N}_*$,
$\| \rho \|_{{\cal N}_*}$
$=1$)
converges to
$\rho_0 (\in  {\frak S}^p  ({}{\cal A}^*{}))$
in the sense of weak$^*$ topology of ${\cal A}^*$,
that is,
\begin{align}
\rho(G) 
\xrightarrow[\quad]{} \rho_0 (G)
\;\;
(\forall G \in {\cal A} (\subseteq {\cal N} )
),
\label{eq4}
\end{align}
then
$\rho(F)$
converges to
$\alpha $.
\end{itemize}
And the value of $\rho_0(F)$ is defined by the $\alpha$.

According to the noted idea
\cite{Davi},
an {\it observable}
${\mathsf O}{\; :=}(X, {\cal F},$
$F)$ in 
${{\cal N}}$
is defined as follows:
\par
\par
\begin{itemize}
\item[(i)]
[$\sigma$-field]
$X$ is a set,
${\cal F}
(\subseteq {\cal P}(X)$,
the power set of $X$)
is a $\sigma$-field of $X$,
that is,
{\lq\lq}$\Xi_1, \Xi_2,... \in {\cal F}\Rightarrow \cup_{n=1}^\infty \Xi_n \in {\cal F}$",
{\lq\lq}$\Xi  \in {\cal F}\Rightarrow X \setminus \Xi \in {\cal F}$".
\item[(ii)]
[Countable additivity]
$F$ is a mapping from ${\cal F}$ to ${{\cal N}}$ 
satisfying:
(a):
for every $\Xi \in {\cal F}$, $F(\Xi)$ is a non-negative element in 
${{\cal N}}$
such that $0 \le F(\Xi) $
$\le I$, 
(b):
$F(\emptyset) = 0$ and 
$F(X) = I$,
where
$0$ and $I$ is the $0$-element and the identity
in ${\cal N}$
respectively.
(c):
for any countable decomposition $\{\Xi_1,\Xi_2,\dots ,\Xi_n, ... \}$ 
of 
$\Xi$ $\big($i.e.,
$\Xi,\Xi_n \in {\cal F}\;(n=1,2,3,...)$, 
$\cup_{n=1}^\infty \Xi_n = \Xi$, 
$\Xi_i \cap \Xi_j = \emptyset$ $(i \ne j) \big)$, it holds that
$
F(\Xi)
$
$
=
$
$
\sum_{n=1}^\infty  F(\Xi_n)
$
in the sense of weak$^*$ topology in ${\cal N}$.
\end{itemize}

\par
\vskip0.2cm
\par
\rm
With any {\it system} $S$, a 
basic structure
$[{\cal A},{\cal N}]_{B(H)}$
%
can be associated in which the 
measurement theory (F) of that system can be formulated.
A {\it state} of the system $S$ is represented by an element
$\rho (\in {\frak S}^p  ({}{\cal A}^*{}))$
and an {\it observable} is represented by an observable 
${\mathsf{O}}{\; :=} (X, {\cal F}, F)$ in ${{\cal N}}$.
Also, the {\it measurement of the observable ${\mathsf{O}}$ for the system 
$S$ with the state $\rho$}
is denoted by 
${\mathsf{M}}_{\cal N} ({\mathsf{O}}, S_{[\rho]})$
$\big($
or more precisely,
${\mathsf{M}}_{\cal N} ({\mathsf{O}}{\; :=} (X, {\cal F}, F),$
$ S_{[\rho]})$
$\big)$.
An observer can obtain a measured value $x $
($\in X$) by the measurement 
${\mathsf{M}}_{\cal N} ({\mathsf{O}}, S_{[\rho]})$.
\par
\noindent
\par
The Axiom 1 presented below is 
a kind of mathematical generalization of Born's probabilistic interpretation of quantum mechanics (E).
And thus, it is a statement without reality.

\par
Now we can present Axiom 1 in the $W^*$-algebraic formulation as follows.
\par
\noindent
\bf
Axiom 1
\rm
[
Measurement
].
\it
The probability that a measured value $x$
$( \in X)$ obtained by the measurement 
${\mathsf{M}}_{\cal N} ({\mathsf{O}}{\; :=} (X, {\cal F}, F),$
$ S_{[\rho_0]})$
%
belongs to a set 
$\Xi (\in {\cal F})$ is given by
$
\rho_0( F(\Xi) )
$
if
$F(\Xi)$
is essentially continuous at $\rho_0 ( \in {\frak S}^p  ({}{\cal A}^*{}) )$.
\rm
\par
\vskip0.3cm

\par
Next, we explain Axiom 2.
Let
$[{\cal A}_1,{\cal N}_1]_{B(H_1)}$
and
$[{\cal A}_2,{\cal N}_2]_{B(H_2)}$
be basic structures.
A continuous
linear operator
$\Phi_{1,2}$
$:{{{{\cal N}_2}}}$
(with weak$^*$ topology)
$\to {{{{\cal N}_1}}}$(with weak$^*$ topology)
is called a
\it
Markov operator,
\rm
if
it satisfies that
(i):
$\Phi_{1,2} (F_2) \ge 0$
for any non-negative element
$F_2$ in
${{{{\cal N}_2}}}$,
(ii):
$ \Phi_{1,2}({}I_2{}) = I_1 $,
where $I_k$
is the identity in ${\cal N}_k$,
$(k=1,2)$.
In addition to the above (i) and (ii),
in this paper
we
assume that
$\Phi_{1,2}({\cal A}_2) \subseteq {\cal A}_1$
and
$\sup \{ \|\Phi_{1,2}( F_2) \|_{{\cal A}_1}
\; |
\;
F_2 \in {\cal A}_2
\text{ such that }
\|F_2\|_{{\cal A}_2} \le 1
\}
=1$.

It is clear that
the dual operator
$ \Phi_{1,2}^*{}: $
$
{\cal A}_{1}^*
\to {\cal A}_{2}^*
$
satisfies that
$ \Phi_{1,2}^*{}
(
{\frak S}^m  ({\cal A}_{1}^*)
)
\subseteq
{\frak S}^m  ({\cal A}_{2}^*)
$.

Here note that,
for any observable
${\mathsf{O}}_2{\; :=}({}X , {\cal F}, {F}_2{})$
in ${{{{\cal N}_2}}}$,
the $({}X , {\cal F}, $
$\Phi_{1,2} F_2 )$
is an observable in ${{{\cal N}_1}}$.

%


\par
\vskip0.3cm

\par
\par
Let $(T,\le)$ be a tree, i.e., a partial ordered 
set such that {\lq\lq$t_1 \le t_3$ and $t_2 \le t_3$\rq\rq} implies {\lq\lq$t_1 \le t_2$ or $t_2 \le t_1$\rq\rq}\!.
Put $T^2_\le = \{ (t_1,t_2) \in T^2{}\;|\; t_1 \le t_2 \}$.
Here, note that $T$ is not necessarily finite.

Assume the completeness of the ordered set $T$.
That is,
for any subset $T'( \subseteq T)$
bounded from below
(i.e.,
there exists $t' (\in T)$
such that $t' \le t$
$( \forall t \in T' )$),
there uniquely exists
an element
$\text{inf} (T')$
$\in T$
satisfying the following conditions,
(i):
${\text{inf}} ( T' ) {{\; \leqq \;}}t \; ( \forall t \in T' )$,
(ii):
if
$s {{\; \leqq \;}}t \; \; ( \forall t \in T' )$,
then
$s {{\; \leqq \;}}{\text{inf}} ( T' )$.

\par
\noindent
\par
\noindent
\par
The family
$\{ \Phi_{t_1,t_2}{}: $
${\cal N}_{t_2} \to {\cal N}_{t_1} \}_{(t_1,t_2) \in T^2_\le}$
is called a {\it Markov relation}
({\it due to the Heisenberg picture}),
\rm
if it satisfies the following conditions {\rm (i) and (ii)}.
\begin{itemize}
\item[{\rm (i)}]
With each
$t \in T$,
a basic structure
$[{\cal A}_t,{\cal N}_t]_{B(H_t)}$
is associated.
\item[{\rm (ii)}]
For every $(t_1,t_2) \in T_{\le}^2$, a Markov operator 
$\Phi_{t_1,t_2}{}: {\cal N}_{t_2} \to {\cal N}_{t_1}$ 
is defined.
And it satisfies that
$\Phi_{t_1,t_2} \Phi_{t_2,t_3} = \Phi_{t_1,t_3}$ 
holds for any $(t_1,t_2)$, $(t_2,t_3)$
$ \in$
$ T_\le^2$.
\end{itemize}
\noindent

%
%
When
$ \Phi_{t_1,t_2}^*{}$
$
(
{\frak S}^p  ({\cal A}_{t_1}^*)
)$
$\subseteq
$
$
(
{\frak S}^p  ({\cal A}_{t_2}^*)
)$
holds for any
$
{(t_1,t_2) \in T^2_\le}$,
the Markov relation is said to be
deterministic.
Note that
the classical deterministic Markov relation
is represented by
$\{ \phi_{t_1,t_2}{}: $
${\Omega}_{t_1} \to {\Omega}_{t_2} \}_{(t_1,t_2) \in T^2_\le}$,
where
the continuous map
$\phi_{t_1,t_2}{}: $
${\Omega}_{t_1} \to {\Omega}_{t_2}$
is defined by
\begin{align*}
\Phi_{t_1,t_2}^* (\delta_{\omega_1} )
=
\delta_{
\phi_{t_1,t_2} (\omega_1)}
\quad
(\forall \omega_1 \in \Omega_1)
\end{align*}

\par
\par
\rm
Now Axiom 2 is presented
as follows:
\rm
\par
\noindent
\bf
Axiom 2
\rm
[Causality].
\it
The causality is represented by
a Markov relation 
$\{ \Phi_{t_1,t_2}{}: $
${\cal N}_{t_2} \to {\cal N}_{t_1} \}_{(t_1,t_2) \in T^2_\le}$.

\rm

\par
\vskip2.0cm

$
\text{----------------------------------------------------------------------------------------------------------------------------------}$
%
\par
\noindent
\begin{picture}(500,170)
\thicklines
\put(0,70){
$
\fbox{{\shortstack[l]{world \\ 
view}}}
$
$
\left\{\begin{array}{l}
\!\!\!
\textcircled{\scriptsize R}:
\underset{\text{\scriptsize (realism)}}{\fbox{\text{Aristotle}}}
{\xrightarrow[]{\textcircled{\scriptsize 1}}}
\overset{\text{\scriptsize (monism)}}{\underset{\text{\scriptsize (realism)}}
{\fbox{\text{Newton}}}}
\xrightarrow[]{}
\left\{\begin{array}{llll}
\fbox{\shortstack[l]{theory of \\ relativity}}
\xrightarrow[]{\qquad \qquad \quad \qquad \qquad}
\\
\\
{\fbox{\shortstack[l]{quantum \\ mechanics}}}
{
\xrightarrow[]{}
\left\{\begin{array}{lll}
\xrightarrow[\quad \text{\scriptsize \rm realistic view}\;\;]{{\text{\rm }}}
\\
\\
\xrightarrow[{{\textcircled{\scriptsize 3}}\rm :\text{linguistic turn}}]{
{\textcircled{\scriptsize 2}}\rm:
{dualism}}
\end{array}\right.
}
\end{array}\right.
\\
\\
\!\!\!
\textcircled{\scriptsize L}:
\overset{\text{\scriptsize }}{
\underset{\text{\scriptsize (idealism)}}{\fbox{\shortstack[l]{Plato \\ 
Parmenides}}}
}
\xrightarrow[]{{\textcircled{\scriptsize 4}}}
\overset{\text{\scriptsize (dualism)}}{
\underset{\text{\scriptsize (idealism)}}{\fbox{
{\shortstack[l]{Kant \\ Descartes}}
}}
}
{\xrightarrow[]{\textcircled{\scriptsize 5}}}
\underset{\text{\scriptsize (linguistic view)}}{\fbox{
\shortstack[l]{philosophy \\ of language}
}}
\xrightarrow[{{\textcircled{\scriptsize 7}}:\text{linguistic view}}]{{
{\textcircled{\scriptsize 6}}:
\rm
axiomatization}}
\end{array}\right.
$
}
\put(332,110){
$
\left.\begin{array}{llll}
\; 
\\
\; 
\\
\; 
\\
\;
\end{array}\right\}
\xrightarrow[]{}
\overset{\text{\scriptsize (unsolved)}}{
\underset{\text{\scriptsize (physics)}}{
\fbox{\shortstack[l]{theory of \\ everything}}
}
}
$
}
\put(332,40){
$
\left.\begin{array}{lllll}
\; 
\\
\; 
\\
\; 
\\
\;
\\
\;
\end{array}\right\}
{\xrightarrow[]{\textcircled{\scriptsize 8}}}
\underset{\text{\scriptsize (scientific language)}}{
\fbox{\shortstack[l]{measurement \\ theory (=MT)}}
}
$
}
\put(2,-20){
{\bf Figure 1. 
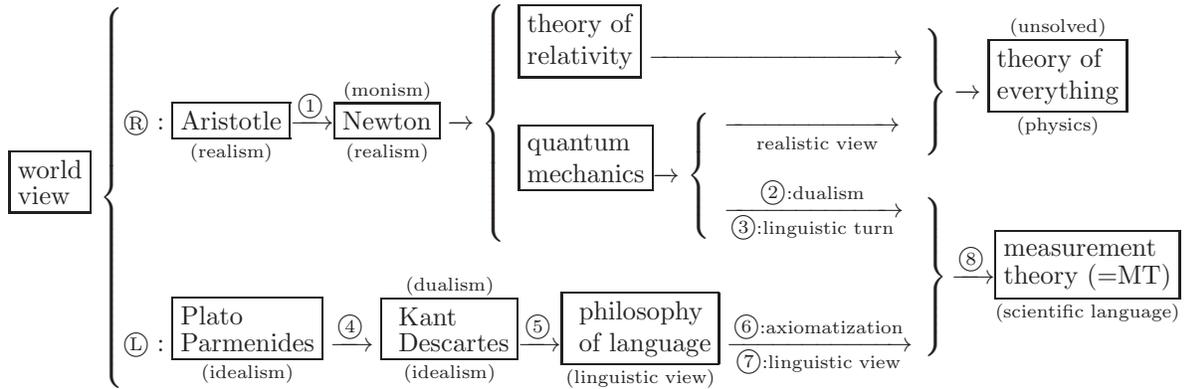
The development of the world views.
For the explanations of
\textcircled{\scriptsize 1}-\textcircled{\scriptsize 8},
see {{{}}}{\cite{Ishi7,Ishi9}}.
}
}
\put(65,-32){
}
\end{picture}

%
\rm
\par
\par
\newpage
\par
\par
\noindent
{\bf
2.3. The Linguistic Interpretation
}

\par
\noindent
\vskip0.2cm
\par
According to
{{{}}}{\cite{Ishi6,Ishi7, Ishi9}},
we shall explain the linguistic interpretation of quantum mechanics.
The measurement theory {{{}}}{(
{F})} asserts
\begin{itemize}
\item[{{{}}}{
{(I)}}]
Obey
Axioms 1 and 2.
And,
describe any ordinary phenomenon according to
Axioms 1 and 2
(in spite that
Axioms 1 and 2 can not be tested experimentally).
\end{itemize}
Still,
most readers 
may be perplexed how to use Axioms 1 and 2
since there are various usages.
Thus, the following problem is significant.
\begin{itemize}
\item[{{{}}}{
{(J)}}]
How should Axioms 1 and 2
be used?
\end{itemize}
Note that reality is not reliable
%
since Axioms 1 and 2 are statements without reality.

Here, in spite of the linguistic turn
(
{\bf Figure 1}:\textcircled{\scriptsize 3})
and the mathematical generalization
from
$B(H)$ to
a $C^*$-algebra
${\cal A}$,
we consider that
the dualism (i.e.,
the main spirit of so called Copenhagen interpretation)
of quantum mechanics
is
inherited
to measurement theory.
Thus, we present the following interpretation
{{{}}}{
{(K)}}
[={{{}}}{{{{}}}{(
{K$_1$})}--{{{}}}{(
{K$_3$})}}].
$\;$
That is,
as the answer to the
\par
\noindent
question
{{{}}}{
{(J)}}, we propose:
\begin{itemize}
\item[{{{}}}{
{(K$_1$)}}]
Consider the dualism composed of {\lq\lq}observer{\rq\rq} 
\newpage
and {\lq\lq}system( =measuring object){\rq\rq}.
And therefore,
{\lq\lq}observer{\rq\rq} and {\lq\lq}system{\rq\rq}
must be absolutely separated.
See {\bf Figure 2}.
\item[{{{}}}{
{(K$_2$)}}]
Only one measurement is permitted.
And thus,
the state after a measurement
is meaningless
$\;$
since it 
can not be measured any longer.
Also, the causality should be assumed only in the side of system,
however,
a state never moves.
Thus,
the Heisenberg picture should be adopted
rather than the Schr\"{o}dinger picture.
\item[{{{}}}{
{(K$_3$)}}]
Also, the observer
does not have
the space-time.
Thus, 
the question:
{\lq\lq}When and where is a measured value obtained?{\rq\rq}
is out of measurement theory,
\end{itemize}
\par
\noindent
and so on.

\par
\noindent
\vskip0.2cm
\par
\rm
\par
\noindent
\vskip0.5cm
\noindent
\unitlength=0.5mm
\begin{picture}(200,72)(15,0)
\put(-8,0)
{
\allinethickness{0.2mm}
\drawline[-40](80,0)(80,62)(30,62)(30,0)
\drawline[-40](130,0)(130,62)(175,62)(175,0)
\allinethickness{0.5mm}
\path(20,0)(175,0)
%
\put(14,-5){
\put(37,50){$\bullet$}
}
\put(50,25){\ellipse{17}{25}}
\put(50,44){\ellipse{10}{13}}
\put(0,44){\put(43,30){\sf \footnotesize{observer}}
\put(42,25){\scriptsize{(I(=mind))}}
}
\put(7,7){\path(46,27)(55,20)(58,20)}
\path(48,13)(47,0)(49,0)(50,13)
\path(51,13)(52,0)(54,0)(53,13)
\put(0,26){
\put(142,48){\sf \footnotesize system}
\put(143,43){\scriptsize (matter)}
}
\path(152,0)(152,20)(165,20)(150,50)(135,20)(148,20)(148,0)
\put(10,0){}
\allinethickness{0.2mm}
\put(0,-5){
\put(130,39){\vector(-1,0){60}}
\put(70,43){\vector(1,0){60}}
\put(92,56){\sf \scriptsize \fbox{observable}}
\put(58,50){\sf \scriptsize }
\put(57,53){\sf \scriptsize \fbox{\shortstack[l]{measured \\ value}}}
\put(80,44){\scriptsize \textcircled{\scriptsize a}interfere}
\put(80,33){\scriptsize \textcircled{\scriptsize b}perceive a reaction}
\put(130,56){\sf \scriptsize \fbox{state}}
}
}
\put(30,-15){\bf
{\bf Figure 2}. 
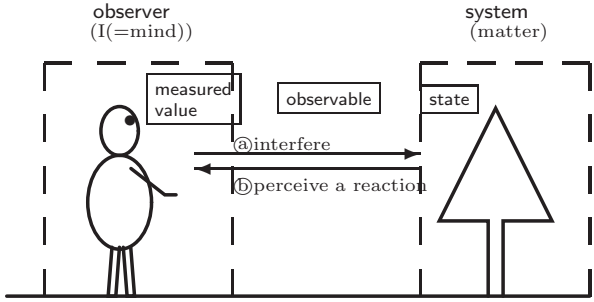
Dualism
in MT.
}
\end{picture}
\vskip1.3cm
\par
\noindent
In this sense,
we consider that
measurement theory holds as a kind of
language-game
(with the rule
(Axioms 1 and 2, Interpretation {{{}}}{
{(K)}}),
and therefore,
measurement theory
is regarded as
the axiomatization (
{\bf Figure 1}:\textcircled{\scriptsize 6}) of
the philosophy of language
(i.e., Saussure's linguistic world view).
For the precise explanations of {\bf Figures 1 and 2},
see {{{}}}{\cite{Ishi7,Ishi9}}.
\par
Note that
quantum mechanics (E) has many interpretations.
On the other hand,
we believe that
the interpretation of measurement theory (F)
is uniquely determined 
as in the above.
This is our main reason to propose the linguistic interpretation
of quantum mechanics.
We believe that this uniqueness is essential to the justification of
Heisenberg's uncertainty principle
({\it cf.}
\cite{IshiU, Ishi9}).

\par
\vskip0.5cm

\par
\par
\par
\noindent
$\bf{Example\;1}$
[Simultaneous measurement]
$\;\;\;$
For each
$k=1,$
$2,...,K$,
consider a measurement
${\mathsf{M}}_{\cal N} ({\mathsf{O}_k}$
${ :=}$
$ (X_k,$
${\cal F}_k,$
$F_k),$
$ S_{[\rho]})$.
$\;$
However,
since
the (K$_2$)
says that
only one measurement is permitted,
the
$\{
{\mathsf{M}}_{\cal N} ({\mathsf{O}_k},$
$S_{[\rho]})
\}_{k=1}^K$
should be reconsidered in what follows.
Under the commutativity condition such that
\begin{align}
&
F_i(\Xi_i) F_j(\Xi_j) 
=
F_j(\Xi_j) F_i(\Xi_i)
\label{eq5}
%
\\
&
\quad
(\forall \Xi_i \in {\cal F}_i,
\forall \Xi_j \in  {\cal F}_j , i \not= j),
\nonumber
\end{align}
we can
define the simultaneous observable
${\text{\large $\times$}}_{k=1}^K {\mathsf{O}_k}$
$=({\text{\large $\times$}}_{k=1}^K X_k ,$
$ \boxtimes_{k=1}^K {\cal F}_k,$
$ 
{\text{\large $\times$}}_{k=1}^K {F}_k)$
in ${\cal N}$ such that
\begin{align}
({\text{\large $\times$}}_{k=1}^K {F}_k)({\text{\large $\times$}}_{k=1}^K {\Xi}_k )
=
F_1(\Xi_1) F_2(\Xi_2) \cdots F_K(\Xi_K)
\label{eq6}
\\
\;
(
\forall \Xi_k \in {\cal F}_k,
\forall k=1,...,K
).
\qquad
\qquad
\nonumber
\end{align}
where
$({\text{\large $\times$}}_{k=1}^K X_k ,$
$ \boxtimes_{k=1}^K {\cal F}_k)$
is the product measurable space.
Then, 
the above
$\{
{\mathsf{M}}_{\cal N} ({\mathsf{O}_k},S_{[\rho]})
\}_{k=1}^K$
is,
under the commutativity condition {{{}}}{(\ref{eq5})},
represented by the simultaneous measurement
${\mathsf{M}}_{{{{\cal N}}}} (
{\text{\large $\times$}}_{k=1}^K {\mathsf{O}_k}$,
$ S_{[\rho]})$.

\par
\noindent
$\bf{Example\;2}$
[How to use Axiom 2 (Causality)]
Consider a finite tree
$(T{\; :=}\{t_0, t_1, ..., t_n \},$
$ \le )$
with the root $t_0$.
This is also characterized by
the map
$\pi: T \setminus \{t_0\} \to T$
such that
$\pi( t)= \max \{ s \in T \;|\; s < t \}$.
Let
$\{ \Phi_{t', t} : {\cal N}_{t}  \to {\cal N}_{t'}  \}_{ (t',t)\in
T_\le^2}$
be a Markov relation,
which is also represented by
$\{ \Phi_{\pi(t), t} : {\cal N}_{t}  \to {\cal N}_{\pi(t)}  \}_{ 
t \in T \setminus \{t_0\}}$.
Let an observable ${\mathsf O}_t{\; :=}
(X_t, {\cal F}_{t}, F_t)$ in the ${\cal N}_t$ 
be given for each $t \in T$.
Consider the pair
${\mathbb O}_T$
$=$
$[\{ {\mathsf O}_t \}_{t\in T},
$
$\{ \Phi_{\pi(t), t} : {\cal N}_{t}  \to {\cal N}_{\pi(t)}  \}_{ 
t \in T \setminus \{t_0\}}$
$]$,
which is called a sequential observable.
Let $\rho_{0} \in {\frak S}^p  ({}{\cal A}_{t_0}^*{})$.
Consider {\lq\lq}measurement" such as
\begin{itemize}
\item[(L)]
a measurement of a sequential  observable
${\mathbb O}_T$
for the system with a
$\rho_0$.
%
\end{itemize}
where the meaning of the (L) is not clear yet.
Recalling that
the (K$_3$) says that a state never moves,
we consider the meaning of the (L) as follows:
For each $s \in T$,
put $T_s =\{ t \in T \;|\; t \ge s\}$.
And define the observable
${\widehat{\mathsf O}}_s
=({\text{\large $\times$}}_{t \in T_s}X_t, \boxtimes_{t \in T_s}{\cal F}_t, {\widehat{F}}_s)$
in ${\cal N}_s$
(due to Heisenberg picture)
as follows:
\par
\noindent
\begin{align}
\widehat{\mathsf O}_s
&=
\left\{\begin{array}{ll}
{\mathsf O}_s
\quad
&
\!\!\!\!\!\!\!\!\!\!\!\!\!\!\!\!\!\!
\text{(if $s \in T \setminus \pi (T) \;${})}
\\
\\
{\mathsf O}_s
{\text{\large $\times$}}
({}\bigtimes_{t \in \pi^{-1} ({}\{ s \}{})} \Phi_{ \pi(t), t} \widehat {\mathsf O}_t{})
\quad
&
\!\!\!\!\!\!
\text{(if $ s \in \pi (T) ${})}
\end{array}\right.
\nonumber
\\
&
\!\!\!\!\!
\label{eq7}
\end{align}
if
the commutativity condition holds
(i.e.,
if the simultaneous observable
${\mathsf O}_s
{\text{\large $\times$}}
({}\bigtimes_{t \in \pi^{-1} ({}\{ s \}{})} \Phi_{ \pi(t), t}
$
$\widehat {\mathsf O}_t{})$
exists)
for each $s \in \pi(T)$.
Using {{{}}}{(\ref{eq7})} iteratively,
we can finally obtain the observable
$\widehat{\mathsf O}_{t_0}$
(which is also denoted by
$\widehat{\mathsf O}_{T}$
)
in ${\cal N}_{t_0}$,
which is regarded as the realization of
${\mathbb O}_T$.
Thus the above (L) is represented by
the measurement
${\mathsf{M}}_{{{\cal N}_{t_0} }} (\widehat{\mathsf{O}}_{T},$
$ S_{[\rho_0]})$.

\par
\vskip0.5cm
\par
\noindent
{\bf 2.4. 
Leibniz-Clarke Correspondence (Space-Time Problem)
}
\par

\par
\noindent
\par
Consider a basic structure
$[{\cal A},{\cal N}]_{B(H)}$.
%
Let
${\cal A}_S$
($\subseteq {\cal N}$)
be the commutative $C^*$-subalgebra.
Note that
${\cal A}_S$
is represented
such that
${\cal A}_S$
$=C_0(\Omega_S)$
for some locally compact Hausdorff space $\Omega_S$
({\it cf.} {{{}}}{{{}}}{{\cite{Saka}}}).
As seen in the formula
{{{}}}{(\ref{eq3})},
the $\Omega_S$ is called a {\it spectrum}.
For example,
consider one particle quantum system,
formulated in a basic structure:
\begin{align*}
[B_c(L^2({\mathbb R}^3)),B(L^2({\mathbb R}^3))]_{B(L^2({\mathbb R}^3))}.
\end{align*}
Then, we can choose
the commutative $C^*$-algebra
$C_0({\mathbb R}^3)$
$( \subset$
$
B(L^2({\mathbb R}^3)
)$,
and thus, we get the spectrum
${\mathbb R}^3$.
This simple example will make us propose {{{}}}{
{the (M$_2$)}}
later.
\par
In
Leibniz-Clarke correspondence
(1715--1716),
they
(i.e.,
Leibniz
and Clarke(=Newton's friend)
)
discussed
{\lq\lq}space-time problem{\rq\rq}.
Their ideas are summarized as follows:
\begin{itemize}
\item[{{{}}}{
{(M$_1$)}}]
$
\!\!
\cases
\!\!\!
\textcircled{\scriptsize R}
\!\!
:
\!\!
\underset{\text{\scriptsize (realistic world view)}}{\text{Newton, Clarke}}
&
\!\!\!\!\!\!\!\!
\cdots
\overset{\text{\scriptsize (space-time in physics)}}{
\underset{\text{\scriptsize }}{
\fbox{\text{realistic space-time}}}
}
\\
\\
\!\!\!
\textcircled{\scriptsize L}
\!\!
:
\!\!\!
\underset{\text{\scriptsize (linguistic world view)}}{\text{Leibniz}}  
&
\!\!\!\!\!\!\!\!
\cdots
\overset{\text{\scriptsize (space-time in language)}}{
\underset{\text{\scriptsize }}{
\fbox{\text{linguistic space-time}}}
}
\endcases
$
\end{itemize}
That is,
Newton considered
{\lq\lq}What is space-time?{\rq\rq}.
On the other hand,
Leibniz considered
{\lq\lq}How should space-time be represented?{\rq\rq},
though he did not propose
his language.
Measurement theory is in Leibniz's side,
and asserts that
\begin{itemize}
\item[{{{}}}{
{(M$_2$)}}]
Space should be described as a kind of spectrum.
And
time should be described as a kind of tree.
In other words,
time is represented by a parameter
$t$
in a linear ordered tree $T$.
\end{itemize}
Therefore, we think that
the Leibniz-Clarke debates
should be essentially regarded as
{\lq\lq}the linguistic world view
\textcircled{\scriptsize L}{\rq\rq}
vs. {\lq\lq}the realistic world view \textcircled{\scriptsize R}{\rq\rq}
in {\bf Figure 1}.
Hence, the statement {{{}}}{
{(M$_2$)}} should be added to
Interpretation
{{{}}}{
{(K)}} as sub-interpretation
of measurement theory.
\par
\noindent
\vskip0.5cm
\par
\noindent
\par
\noindent

\vskip0.3cm


\par
\noindent
{\bf 3. 
\large
Preliminary mathematical results}
\par
\newcommand{\qp}{\overset{{\rm qp}}{\pmb{\pmb\times}}}
%

\par
   Let ${\widehat \Lambda}$ be an index set.
For each $\lambda\in{\widehat \Lambda}$, consider a set $X_\lambda$.
For any subsets
$\Lambda_{1} \subseteq \Lambda_{2}({} \subseteq {\widehat \Lambda}{})$,
$P_{\Lambda_{1},\Lambda_{2}}$ is defined by the natural projection such that:
\begin{align}
\mathop{\mbox{\Large $\times$}}_{\lambda\in\Lambda_{2}}X_{\lambda}
\ni (x_\lambda)_{\lambda \in \Lambda_2}
\xrightarrow[P_{\Lambda_{1},\Lambda_{2}}]{}
(x_\lambda)_{\lambda \in \Lambda_1}
\in
\mathop{\mbox{\Large $\times$}}_{\lambda\in\Lambda_{1}}X_{\lambda}.
\nonumber
\end{align}
For each $\lambda\in{\widehat \Lambda}$, consider an observable
$(X_{\lambda},{\cal F}_{\lambda},F_{\lambda})$ in $W^*$-algebra ${\cal N}$.
And consider the  quasi-product observable
${}{{}\mathsf O}$
$\equiv$
$(${}$\mbox{\Large $\times$}_{\lambda\in{\widehat \Lambda}}X_{\lambda},$
$\mbox{$\bigstimes$}_{\lambda\in{\widehat \Lambda}} {\cal
F}_{\lambda},$
$F_{\widehat \Lambda}${}$)$
of
$\{$
$(X_{\lambda},$
${\cal F}_{\lambda},$ $F_\lambda)$
$\; | \;$ ${\lambda}\in{\widehat \Lambda}$
$\}$,
which
is defined by the observable such that:
\begin{align}
{\widehat F}_{\widehat \Lambda} ({}P_{\{ \lambda \},{\widehat \Lambda}}^{-1} ({}\Xi_{\lambda}{}))
=
F_{\lambda}(\Xi_{\lambda})
\qquad
({}
\forall \Xi_\lambda \in {\cal F}_\lambda,
\forall {\lambda }\in{\widehat \Lambda}{}),
\label{eq8}
\end{align}
though
the existence and the uniqueness of a quasi-product
observable are not guaranteed in general.
The following theorem says something about
the existence and uniqueness of
the quasi-product observable.
\par
\noindent
{\bf Theorem 1}
\rm
[{}$W^*$-algebraic Kolmogorov
extension theorem, {\it cf.} ${{{}}}${{{}}}{\cite{Kolm, Ishi4}}{}].
\it
For each $\lambda\in{\widehat \Lambda}$, consider
   a Borel measurable space
   $({}X_\lambda , {\cal F}_{\lambda}{})$,
   where
   $X_\lambda$
   is
   a separable complete metric space.
Define the set
${\cal P}_0 ({\widehat \Lambda})$
such as
$ {\cal P}_0 ({\widehat \Lambda}) \equiv
\{
\Lambda \subseteq {\widehat \Lambda} \;  | \;
\Lambda ~\mbox{\rm is finite  }
\}
$.
      Assume that the family of
     the observables
     $\bigl\{$
     ${{{}\mathsf O}}_{\Lambda} \equiv$
     $(${}     $\mathop{\mbox{\Large $\times$}}_{\lambda\in\Lambda}X_{\lambda},
$
     $\mathop{\mbox{$\bigstimes$}}_{\lambda\in\Lambda}{\cal
F}_{\lambda},$
     $F_{\Lambda}$
     $)$
     $~|~$
     $\Lambda\in{\cal P}_0 ({\widehat \Lambda})$
     $\bigr\}$
     in ${\cal N}$
     satisfies the following {\lq\lq consistency
condition\rq\rq}$:$
       \begin{itemize}
       \item[{\rm (N)}]
       for any $\Lambda_1$, $\Lambda_2$
$\in$
${\cal P}_0 ({\widehat
\Lambda})$
       such that $\Lambda_1 \subseteq \Lambda_2$,
       \\
       it holds that
       \end{itemize}
\begin{align}
&
F_{\Lambda_2}
\bigl({}
       P_{\Lambda_{1},\Lambda_{2}}^{-1}({\Xi}_{\Lambda_1}{})
       \bigr)
       =
       F_{\Lambda_1}
       \bigl({}{\Xi}_{\Lambda_1} \bigr)
       \nonumber
       \\
       &
       \quad
       ({}\forall {\Xi}_{\Lambda_1} \in
       \mathop{\mbox{$\bigstimes$}}_{\lambda\in\Lambda_1}
       {\cal F}_{\lambda}{}).
\label{eq9} 
\end{align}
\par
\it
\noindent
Then, 
$\;\;$
there uniquely exists the observable
${\widehat{\mathsf O}}_{\widehat{\Lambda}}$
$\equiv$
$\bigl(\mathop{\mbox{\Large $\times$}}_{\lambda\in{\widehat
\Lambda}}X_{\lambda}, $
$\mathop{\mbox{$\bigstimes$}}_{\lambda\in{\widehat \Lambda}}
{\cal
F}_{\lambda},$
${\widehat F}_{\widehat \Lambda}
\bigr)$
in ${\cal N}$ such that:
\begin{align*}
&
       {\widehat F}_{\widehat \Lambda}
       \bigl({}
       P_{\Lambda, {\widehat \Lambda}}^{-1}({\Xi}_{\Lambda}{})
       \bigr)
       =
       F_{\Lambda}  \bigl({}{\Xi}_{\Lambda} \bigr)
       \\
&
       \quad
       ({}\forall {\Xi}_{\Lambda} \in
       \mathop{\mbox{$\bigstimes$}}_{\lambda\in\Lambda}
       {\cal F}_{\lambda},~
       \forall\Lambda\in{\cal P}_0 ({\widehat \Lambda}){}).
\end{align*}
\par 
\par
\noindent
{\it $\;\;\;\;$ Proof.}$\;\;$
\rm
See {{{}}}{\cite{Ishi4}}.
\qed
\noindent
\par 
\vskip0.5cm
\par
As mentioned in \cite{Ishi6},
we believe that
\begin{itemize}
\item[(O)]
the utility of Kolmogorov's extension theorem
is due to
the interpretation
(K$_2$).
\end{itemize}
We think that this view is the most essential in all statements concerning
Kolmogorov's extension theorem.
\rm
\par
\vskip0.3cm
\par
Consider
a Borel measurable space
$({}X , {\cal B}_{X}{})$,
where
$X$
is
a (locally compact)
separable complete metric space.
Let $\Omega$ ba locally compact Hausdorff space
with a suitable measure $\nu$.
Let $g$ be a quantity,
that is,
a continuous map $g: \Omega \to X$.
Define the 
observable
${\sf O}_g =(X, {\cal B}_X , G )$
in $L^\infty ( \Omega, \nu )$
such that
\begin{align}
&
[G(\Xi)](\omega) 
=
\chi_{g^{-1}(\Xi )} (\omega )
=
\cases
1 \;\;&
\text{ if }
\omega \in {g^{-1}(\Xi )}
\\
0
\;\;
&
\text{ if }
\omega \notin {g^{-1}(\Xi )}
\endcases
\nonumber
\\
&
\qquad
\qquad
\qquad
(\forall \Xi \in {\cal B}_X, \omega \in \Omega ).
\label{eq10}
\end{align}

%
\par
\noindent
{\bf Lemma 1}
\rm
[The measurement of a quantity].
\it
\it
Let
${\mathsf O}_g=(X,{\cal B}_{X}, G)$
be the observable
induced by
a quantity
${ g}:\Omega \to {X}$
as in {{{}}}{(\ref{eq10})}.
$\;\;$
Let
$x $
$(\in{X})$
be a measured value obtained by
the measurement
${{\mathsf M}}_{L^\infty (\Omega, \nu)}(
{\mathsf O}_g
,$
$ S_{[\delta_{\omega_0}]} )$.
Then,
we can surely believe
that
$x= { g}({\omega_0})$.
That is,
for any open set
$D$
$(\subseteq X)$
such that
${ g}({\omega_0})
\in$
$D$,
the probability that
a measured value obtained
by
${{\mathsf M}}_{L^\infty (\Omega, \nu)}(
{\mathsf O}_g
, S_{[\delta_{\omega_0}]} )$
belongs to
$D$
is equal to
$1$.

\par
\noindent
{\it $\;\;\;$ Proof.}$\;$
\rm
Let
$D (\in {\cal B}_{X})$
be any open set
such that
${ g}({\omega_0})
\in$
$D$.
According to
Axiom 1,
the probability
that
a measured value
$x$
obtained by
the measurement
${{\mathsf M}}_{L^\infty (\Omega, \nu)}(
{\mathsf O}_g
, S_{[\delta_{\omega_0}]} )$
belongs to
$D$
is given by
$
\chi_{_{{ g}^{-1}(D)}} ({\omega_0} )=1$.
Since
$D$
is arbitrary,
we can surely believe
that
$x={ g}({\omega_0})$.
\qed

\par
\noindent
$\bf{Lemma\;2}$.
\it
Consider a finite tree
$(T{\; :=}\{t_0, t_1, ..., t_n \},$
$ \le )$
with the root $t_0$.
Let
$\{ \Phi_{t', t} : L^\infty (\Omega_{t}) \to $
$
L^\infty (\Omega_{t'}) 
$
$
\}_{ (t',t)\in
T_\le^2}$
be a deterministic Markov relation,
which is also represented by
the deterministic maps
$\{ \phi_{\pi(t), t} : 
\Omega_{\pi(t)} \to 
\Omega_{t}
\}_{T\setminus \{t_0\}}$.
For each $t \in T$,
consider a continuous map $g_t : \Omega_t \to X_t$.
and consider an observable ${\mathsf O}_{g_t}$
${\; :=}
$
$
(X_t,$
$ {\cal B}_{X_t},$
$ G_t)$ in the $
L^\infty (\Omega_{t})
$
induced by a quantity $g_t$.
Let
$\widehat{\mathsf O}_T$
be
the realization of the deterministic sequential observable
${\mathbb O}_T$
$=$
$[\{ {\mathsf O}_{g_t} \}_{t\in T},
$
$
\{ \phi_{\pi(t), t} : 
\Omega_{\pi(t)} \to$
$
\Omega_{t}
\}_{T\setminus \{t_0\}}
]$.
Let
$(x_t )_{t \in T}$
$(\in \bigtimes_{t \in T} X_t)$
be a measured value obtained by
the measurement
${{\mathsf M}}_{L^\infty (\Omega, \nu)}$
$(
\widehat{\mathsf O}_T
, S_{[\delta_{\omega_0}]} )$.
$\;$
Then,
we can surely believe
that
$x_{t} = {g_t}(\phi_{0,t} ({\omega_0}))$
$\;(\forall t \in T)$.
That is,
for any open set
$D_t$
$(\subseteq X_t)$
such that
${ g_t}(\phi_{0,t}({\omega_0}))
\in$
$D_t$
$\;$
$(\forall t \in T)$,
the probability that
a measured value $(x_t)_{t \in T} $
obtained
by
${{\mathsf M}}_{L^\infty (\Omega_{t_0}, \nu_{t_0} )}(
\widehat{\mathsf O}_T
, S_{[\delta_{\omega_0}]} )$
belongs to
$\bigtimes_{t \in T} D_t $
is equal to
$1$.
\rm
\par
\noindent

\par
\noindent
{\it $\;\;\;\;$ Proof.}$\;\;$
\rm
This is a slight generalization of Lemma 1.
Thus, the proof is easy as follows.
For each $t \in T$, consider any open set
$D_t (\in {\cal B}_{X_t})$
such that
${ g_t}(\phi_{0, t} ({\omega_0}))$
$(= (g_t \circ \phi_{0,t})(\omega_0)
)$
$
\in$
$D_t$.
According to
Axiom 1,
the probability
that
a measured value
$(x_t)_{t \in T}$
obtained by
the measurement
${{\mathsf M}}_{L^\infty (\Omega_{t_0}, \nu_{t_0})}(
\widehat{\mathsf O}_T
, S_{[\delta_{\omega_0}]} )$
belongs to
$\bigtimes_{t \in T} D_t$
is given by
$
\Pi_{{}_{t\in T}} \chi_{_{{ (g_t \circ \phi_{0,t})}^{-1}(D_t)}} ({\omega_0} )=1$.
This completes the proof.
\qed

\par

\par
\vskip0.5cm

\noindent
{\bf \large
4. 
Zeno's paradox in measurement theory}
\par
\par
\noindent
\par
\noindent
\par
\noindent
{
\begin{itemize}
\item[{\bf 4.1.}]
{
\bf
What is Zeno's paradox?}
\end{itemize}
}
\par
\noindent

Now we can review the question:{\lq\lq}What is Zeno's paradox?"
in Section 1.4 as follows:
\begin{itemize}
\item[(P$_1$)]
From {\bf Figure 1},
choose the proper world view
on which
the formula
{{{}}}{(\ref{eq1})}
should be based!
\item[(P$_2$)]
Or, if there is no proper world view
in this figure,
propose and add a new world view
to 
{\bf Figure 1}.
\end{itemize}
Of course, we shall execute the our assertion (D)
in the following section.
\par

\par
\noindent
\par
\noindent
\par
\noindent
{
\begin{itemize}
\item[{\bf 4.2.}]
{
\bf
Zeno's paradox in measurement theory}
\end{itemize}
}
\par
\noindent

\par
According to
the space-time problem:
(M$_2$),
define the time axis
by
${\widehat T}=[0, \infty)$
with the usual order $\le$.
For each
$t \in$
${\widehat T}$,
consider a classical basic structure
$[C_0 (\Omega_t ),$ 
$L^\infty (\Omega_t, \nu_t )]_{B(H)}$.
Here it may be usual to assume that
$\Omega_t = \Omega $
$(\forall t)$.
Further, consider a quantity
$g_t : \Omega_t \to X_t (\equiv {\mathbb R}^2 )$,
which,
by {{{}}}{(\ref{eq10})}, induces the observable
${\mathsf O}_{g_t}=( X_t, {\cal B}_{X_t}, G_t )$
in $L^\infty (\Omega_t )$,
and consider the
deterministic sequential observable
${\mathbb O}_{\widehat T}$
$=$
$[\{ {\mathsf O}_{g_t} \}_{t\in {\widehat T}},
$
$
\{ \phi_{t', t} : 
\Omega_{t'} \to$
$
\Omega_{t}
\}_{(t', t)  \in {{\widehat T}_\le^2} }
]$.
Here,
for any finite $T ( \in {\cal P}_0({\widehat T}))$,
we have
a
deterministic sequential observable
${\mathbb O}_{T}$
$=$
$[\{ {\mathsf O}_{g_t} \}_{t\in { T}},
$
$
\{ \phi_{t', t} : 
\Omega_{t'} \to$
$
\Omega_{t}
\}_{(t', t)  \in {T_\le^2} }
]$,
which has the realization
$\widehat{\mathsf O}_T$
in $L^\infty ( \Omega_0)$
({\it cf.} Lemma 2).
Consider the family
$\{ \widehat{\mathsf O}_T \;|\; T
 \in {\cal P}_0({\widehat T}) \}$,
which clearly satisfies the consistency condition (N).
Thus, by Theorem 1, we get
the realization
$\widehat{\mathsf O}_{\widehat T}$
in $L^\infty ( \Omega_0)$
of
the sequential observable
${\mathbb O}_{\widehat T}$.
Let
$(x_t )_{t \in {\widehat T}}$
$(\in \bigtimes_{t \in {\widehat T}} X_t)$
be a measured value obtained by
the measurement
${{\mathsf M}}_{L^\infty (\Omega_0, \nu_0)}(
\widehat{\mathsf O}_{\widehat T}
, S_{[\delta_{\omega_0}]} )$.
Then,
we can surely believe
that
\begin{align}
x_{t} = {g_t}(\phi_{0,t} ({\omega_0})
\quad
(\forall t \in {\widehat T})
\label{eq11}
\end{align}
That is,
for any 
$T$
$ \in {\cal P}_0({\widehat T})$
and
any
open set
$D_t$
$(\subseteq X)$
such that
${ g_t}(\phi_{0,t}({\omega_0}))
\in$
$D_t$
$\;$
$(\forall t \in \forall T )$,
the probability that
a measured value $(x_t)_{t \in {\widehat T}} $
obtained
by
${{\mathsf M}}_{L^\infty (\Omega_{t_0}, \nu_{t_0} )}(
\widehat{\mathsf O}_{\widehat T}
, S_{[\delta_{\omega_0}]} )$
belongs to
$P_{T, \widehat T}^{-1} (\bigtimes_{t \in T} D_t )$
is equal to
$1$.

Recall the formula {{{}}}{(\ref{eq1})}.
If we put
\begin{itemize}
\item[(Q)]
$x_t = (\xi (t), \eta (t))
=(vt, \gamma vt + a )$
$\quad$
$(\forall t \in [0, \infty ))$,
\end{itemize}
we can describe the formula
{{{}}}{(\ref{eq1})} in terms of measurement theory.
That is,
in measurement theory,
the formula {{{}}}{(\ref{eq1})}
should be understood as the equation of measured values.
Of course, the calculation of the (Q) is the same as that of
Section 1.3.
\par
\noindent
{\bf Remark 1}.
(i):
In {{{}}}{\cite{Ishi5}}, the above was discussed in the special case that
$\Omega_t = X_t = {\mathbb R}^2$.
Now we think that
it is not sufficient
and it should be improved as mentioned in this section.
However, in the sense mentioned in the abstract,
we believe that
Zeno's paradoxes were already solved in measurement theory
{{{}}}{\cite{Ishi1,Ishi2,Ishi3,Ishi4,Ishi5,Ishi6,Ishi7,Ishi8,Ishi9,Ishi10,Ishi11}}.
\\
(ii):
The beginners of philosophy may misunderstand
{\lq\lq}Achilles and the tortoise"
as the elementary mathematical problem concerning
infinite series.
In order to avoid the confusion, we choose
{\lq\lq}Achilles and the tortoise"
and not
{\lq\lq}flying arrow paradox",
though the latter is the most excellent
in all Zeno's paradoxes.

\par
\vskip0.5cm

\noindent
{\bf \large
5. 
Conclusions}
\par
\par

\par
\noindent
\par

What we executed in this paper
is
merely
the translation from 
{\lq\lq}ordinary language"
to
{\lq\lq}scientific language",
that is,
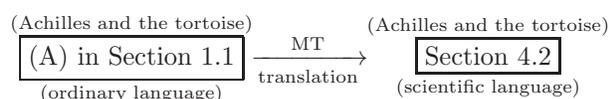
\begin{align*}
\overset{\text{\scriptsize (Achilles and the tortoise)}}{
\underset{\text{\scriptsize (ordinary language)}}{
\fbox{\text{(A) in Section 1.1}}}
}
\!\!
\xrightarrow[\text{translation}]{\text{MT}}
\!\!
\overset{\text{\scriptsize (Achilles and the tortoise)}}{
\underset{\text{\scriptsize (scientific language)}}{
\fbox{\text{Section 4.2}}}
}
\end{align*}
We believe that
this translation is just {\lq\lq}the mechanical world view"
or
{\lq\lq}the method of science"
(at least, science in the series
\textcircled{\scriptsize L}
of
{\bf Figure 1}).
That is,
ordinary science
(at least, its basic statements )
should be described in terms of measurement theory.
For example,
for the translation of equilibrium statistical mechanics
and the Monty-Hall problem,
see {{{}}}{\cite{Ishi10}}and
\cite{Ishi11} respectively
.

\noindent
\par
Since Zeno's paradoxes
have the long history of
2500 years,
we should refrain from the immediate conclusion. 
However,
we believe that
our view (C)
is the central subject of
Zeno's paradoxes.

We hope that 
some readers will propose another powerful
scientific language
(as mentioned in (P$_2$)
),
and also,
our assertion will be examined 
from the various points of view.

%
%
\rm
\vskip-0.5cm
\par
\renewcommand{\refname}{
\large 
6. References}
{
\small

\normalsize
}

%

\par
\noindent


\end{document}